\documentclass[reprint,prx,twocolumn,showpacs,amsmath,superscriptaddress,footnoteemail,longbibliography]{revtex4-1}
\usepackage{epsfig}
\usepackage{xcolor}

\renewcommand{\Im}{\operatorname{Im}}

\usepackage{soul}
\newcommand{\crm}[1]{{\st{#1}}}
\newcommand{\cnew}[1]{{\color{red} #1}}

\renewcommand{\crm}[1]{}
\renewcommand{\cnew}[1]{{#1}}

\usepackage[colorlinks=true,citecolor=blue]{hyperref}
\begin{document}
\title{On the density of states of circular graphene quantum dots}

\author{H. Chau Nguyen} 
\email{chau@pks.mpg.de}
\affiliation{Max-Planck-Institut f\"ur Physik komplexer Systeme,  N\"othnitzer Stra{\ss}e 38, D-01187 Dresden, Germany}

\author{Nhung T. T. Nguyen}
\affiliation{Institute of Physics, Vietnam Academy of Science and Technology, 10 Dao Tan, Ba Dinh Distr., 118011 Hanoi, Vietnam}
\affiliation{Graduate University of Science and Technology, Vietnam Academy of Science and Technology, 18 Hoang Quoc Viet, Cau Giay Distr., 122121 Hanoi, Vietnam}

\author{V. Lien Nguyen}
\affiliation{Institute of Physics, Vietnam Academy of Science and Technology, 10 Dao Tan, Ba Dinh Distr., 118011 Hanoi, Vietnam} 
\affiliation{Institute for Bio-Medical Physics, 109A Pasteur, 1st Distr., 710115 Hochiminh City, Vietnam}

\vspace*{1cm}
\begin{abstract}
We suggest a simple approach to calculate the local density of states that effectively applies to any structure created by an axially symmetric potential on a continuous graphene sheet such as circular graphene quantum dots or rings. Calculations performed for the graphene quantum dot studied in a recent scanning tunneling microscopy  measurement [{\sl Gutierrez et al. Nat. Phys. \textbf{12}, 1069--1075 (2016)}] show an excellent experimental-theoretical agreement.
\end{abstract}

\maketitle
Quantum dots are among the most intensively studied nano-structures. From the application point of view, it is desirable to create quantum dots by feasible and controllable confinement potentials. For conventional semiconductors, such a confinement potential can be easily realized experimentally, e.g., using an appropriate system of gates. The gate-induced electrostatic potentials can be tuned externally to confine electrons to localized states with some desired properties~\cite{chakraborty1999quantum}. As for the mono-layer graphene, due to the Klein tunneling, it has been  challenging to experimentally realize the potentials that can induce strictly localized electronic states~\cite{katsnelson2006chiral,neto2009electronic}. 
Fortunately, though electrostatic potentials fail to create truly bound electronic states, they can trap the charge carriers in quasi-bound states (QBSs) with a trapping time long enough to satisfy application requirements~\cite{matulis2008quasibound}. Thus, various confinement potential models have been probed to seek for appropriate QBS-structures~\cite{matulis2008quasibound,chen2007fock,hewageegana2008electron,recher2009bound,bardarson2009electrostatic,downing2011zero,schneider2014density,wu2014scattering,Givaras2009a,Masir2011a,Givaras2012a,kim2012resonant,Rubio2015a}. Notably, most of the potentials probed~\cite{matulis2008quasibound,chen2007fock,hewageegana2008electron,recher2009bound,bardarson2009electrostatic,downing2011zero,schneider2014density,wu2014scattering,Givaras2009a,Masir2011a,Givaras2012a,kim2012resonant,Rubio2015a} are axially symmetric, implying that the examined graphene quantum dots are circular in shape {(circular graphene quantum dots - CGQDs)}.

Each QBS is characterized by its energy and trapping time, expressing respectively as the energy and the width of a resonance emerging in the local density of states (LDOS). One can therefore identify QBSs by analyzing the structure of the LDOS~\cite{matulis2008quasibound}. Alternatively, one can also directly find the energy spectrum of QBSs by solving the Dirac equation with an outgoing wave boundary condition. Generally, the QBS spectrum is then complex: while the real parts give the energy positions of QBSs, the imaginary parts give the inverse of their trapping times~\cite{chen2007fock,hewageegana2008electron,nguyen2009quasi}. 

The interest in CGQDs has particularly raised in the recent tunneling spectroscopy measurements~\cite{zhao2015creating,freitag2016electrostatically,gutierrez2016imaging,lee2016imaging}. It was suggested that the tip of a scanning tunneling microscopy (STM) can be finely adjusted to create a quantum dot on {the} continuous graphene sheet~\cite{zhao2015creating,freitag2016electrostatically,lee2016imaging}. It was claimed that all the graphene quantum dots realized in these experiments are practically circular~\cite{zhao2015creating,freitag2016electrostatically,lee2016imaging}. Impressively, STM is also the tool to detect the LDOS of graphene quantum dots with high precision. In fact, it has been used to explore the electron whispering-gallery mode resonators~\cite{zhao2015creating} and to directly image the wave functions of QBSs~\cite{zhao2015creating,gutierrez2016imaging,lee2016imaging}.



To theoretically describe the aforementioned experimental data, one has to calculate the LDOS for the CGQD of interest. 
In Refs.~\cite{gutierrez2016imaging} and ~\cite{lee2016imaging} the LDOS was calculated using the scattering and the the finite different methods, respectively. 
It was claimed in Ref.~\cite{gutierrez2016imaging} that the experimental data agree well with the calculated LDOS, except for the QBS of lowest angular momentum. This QBS has made a puzzle by experimentally appearing at the energy considerably higher than theoretically predicted.  

In the present work, stimulated by the beautiful STM measurements, we suggest an approach to efficiently calculate the LDOS of any {realistic} CGQD created by an axially symmetric electrostatic potential, avoiding the indirect calculation of scattering coefficients or the computationally expensive finite difference method. As illustrations, we calculate the LDOS in two typical cases of step and smooth confinement potentials. In the former case, the LDOS was calculated for the CGQD measured/calculated in Ref.~\cite{gutierrez2016imaging}. Our results describe very well the whole experimental QBS spectrum reported in Ref.~\cite{gutierrez2016imaging}. Notably, although not as high as the observed value, our results suggest that the QBS of lowest angular momentum is actually expected to be at {the energy higher than that calculated in Ref.~\cite{gutierrez2016imaging}}; some part of the puzzle is therefore {resolved}. The factor that makes our approach different from that used in Ref.~\cite{gutierrez2016imaging} is clarified. For the studied CGQDs, we also show that the resonance widths (RWs) extracted from our calculated LDOS practically coincide with those obtained from the corresponding complex QBS spectrum of the Dirac equation and qualitatively describe the experimental data. In the case of smooth potentials, we calculate the LDOS for CGQDs created by the Lorentzian shape potential, which is believed to describe the potential induced by a charged STM-tip~\cite{downing2011zero}. 

For a general CGQD, the Hamiltonian that describes the low energy properties of trapped electrons has the Dirac-Weyl form:
\begin{equation} 
\mathcal{H} =  \vec{\sigma} \cdot \vec{p} + U(r),
\label{eq:hamilton}
\end{equation}
where $\vec{\sigma} = (\sigma_x , \sigma_y )$ are Pauli matrices, $\vec{p} = - i (\partial_x , \partial_y )$ is the $2$-dimensional momentum operator, and $U(r)$ is an axially symmetric potential. {For the simplicity, we restrict to the case of experiments~\cite{zhao2015creating,freitag2016electrostatically,gutierrez2016imaging,lee2016imaging} where the valley scattering can be neglected and use units such that $\hbar = 1$ and the Fermi velocity $v_F = 1$}.

To calculate the LDOS for the studied CGQDs, one has to solve the eigenvalue equation of Hamiltonian~\eqref{eq:hamilton} with a proper normalization. Suppose $E$ and $\Psi^{(E)}(r,\phi)$ are the associated eigenvalue and eigenfunction of this Hamiltonian. Since the potential $U(r)$ is axially symmetric, the eigenfunction $\Psi^{(E)}(r,\phi)$ can be found in the form 
\begin{equation}
\Psi^{(E)} (r,\phi)= e^{ij \phi} \begin{pmatrix} e^{-i \phi/2}\chi_A^{(E,j)}(r) \\ e^{+i \phi/2} \chi_B^{(E,j)}(r) \end{pmatrix},
\label{eq:j_eigen}
\end{equation} 
where the total angular momentum $j$ takes half-integer values and $\chi_{A/B}^{(E,j)}(r)$ are the radial wave functions on the graphene $A/B$-sublattices. The radial wave function $\chi^{{(E,j)}} (r) = (\chi^{{(E,j)}}_A (r), \chi^{{(E,j)}}_B (r))^T$ follows the equation
\begin{equation}
i \frac{\partial \chi^{{(E,j)}} (r) }{\partial r} = \mathcal{H}_r \chi^{{(E,j)}}(r),
\label{eq:dirac}
\end{equation}
where the formal radial Hamiltonian $\mathcal{H}_r$ is defined by
\begin{equation}
\mathcal{H}_r  = 
\begin{pmatrix}
i \frac{j-\frac{1}{2}}{r} & U(r) - E \\
U(r) - E  & -i \frac{j+\frac{1}{2}}{r} 
\end{pmatrix}.
\end{equation} 

Certainly, because of the circular symmetry of the structure, the LDOS also depends only on the radial coordinate $r$ and can be found as
\begin{equation}
\rho (E,r) = \sum_{j=-\infty}^{+\infty} \rho^{(j)} (E,r),
\label{eq:tldos}
\end{equation}
with
\begin{equation}
\rho^{(j)} (E,r) \propto \frac{1}{\Delta E} \lVert \chi^{{(E,j)}}(r)\rVert ^2,
\label{eq:ldos}
\end{equation}
where $\Delta E$ is the level spacing at the energy $E$ and $\chi^{{(E,j)}}(r)$ has to be subjected to a proper normalization condition. 
{However, for the considered quantum dots, states are only quasi-bound; strictly speaking, the energy spectrum is continuous and the wave function cannot be normalized. } 
To introduce the level spacing $\Delta E$ and the normalization condition for  $\chi^{{(E,j)}}(r)$, we follow the approach suggested in Ref.~\cite{matulis2008quasibound}. In this approach, the quantum dot is {imagined} to be embedded in a {fictitious} large graphene disk of radius ${L}$. This effectively replaces the continuous energy spectrum by dense discrete levels. \cnew{Note that these discrete levels are independent of the local potential applied to the graphene disk to create a quantum dot. The applied potential however changes the wave functions, and thus the electronic density locally. The LDOS describes this perturbation of the electronic density around the quantum dot (relative to the uniform density away from the potential)}. 

As the disk is so large that for much of its area, the potential $U(r)$ is practically flat. Therefore one can assume there exits some distance $r_f  \ll  {L}$ such that for $r \ge r_f$, the potential could be considered constant, $U(r) \equiv U_f$. Consequently, for $r \ge r_f$, the wave function can be expressed in terms of two integral constants $C_f = (C_f^{(1)}, C_f^{(2)} )^T$: 
\begin{equation}
\chi^{{(E,j)}}(r) = W_f (r)  C_{f}, 
\label{eq:C_f}
\end{equation}
where 
\begin{equation}
W_{f} (r)= 
\begin{pmatrix}
J_{j-\frac{1}{2}} (q_{f} r) & Y_{j-\frac{1}{2}} (q_{f} r) \\
i \tau_{f} J_{j+\frac{1}{2}} (q_{f} r) & i \tau_{f} Y_{j+\frac{1}{2}} (q_{f} r)
\end{pmatrix},
\label{eq:chi_f}
\end{equation}
{with $q_{f} = |E-U_{f}| $, $\tau_{f}= \operatorname{sign} (E-U_{f})$, and $J_{j \pm \frac{1}{2}}$ and $Y_{j \pm \frac{1}{2}}$ are Bessel functions of the first and the second kind, respectively}. The two columns of this $W_f (r)$-matrix are just the two independent basic solutions to the radial Hamiltonian $\mathcal{H}_r$ in the region considered (see Ref.~\cite{nguyen2016transfer} for the details). Then, using the fact that the wave function vanishes at $r={L}$, one  finds the level spacing to be~\cite{matulis2008quasibound}
\begin{equation}
\Delta E = \frac{\pi}{{L}}.
\end{equation} 
Next, the normalization condition for the wave function can be found by requiring that the integration of the electronic probability density over the whole disk must be $1$. \cnew{Note that for much of the large disk area outside the quantum dot, the wave function is of the form~\eqref{eq:C_f}. Although the electronic density in this area is small, it spans the whole (fictitious) macroscopic disk. Therefore, the electronic density outside the quantum dot gives the main contribution to the integration taken over the disk, while the contribution from the relatively small area inside the quantum dot can be ignored~\cite{matulis2008quasibound}}. This ultimately results in the following normalization condition~\cite{matulis2008quasibound}:
\begin{equation}
\frac{4 {L} \lVert C_f \rVert ^2}{|E-U_f|}= 1.
\label{eq:normalization_equation}
\end{equation}

The remaining problem is to find an appropriate initial condition so that the differential eq.~\eqref{eq:dirac} can be solved. The case \cnew{when} the potential can be considered to be flat near the origin, namely, $U(r) = U_i $ for $r \le r_i$, has been studied using the $T$-matrix method~\cite{nguyen2016transfer}. In this case, the eigenfunction of eq.~\eqref{eq:dirac} near the origin has the simple form
\begin{equation}
\chi^{{(E,j)}}(r) = \mathcal{N} 
\begin{pmatrix}  
J_{j-\frac{1}{2}} (q_{i} r) \\
i \tau_{i} J_{j+\frac{1}{2}} (q_{i} r) 
\end{pmatrix},
\label{eq:flat_initial}
\end{equation}
with $q_{i} = |E-U_{i}|$, $\tau_{i}= \operatorname{sign} (E-U_{i})$ and $\mathcal{N}$ being the normalization coefficient. One then can just take the solution~\eqref{eq:flat_initial} at $r_i$ as the initial values and solve eq.~\eqref{eq:dirac} for $\chi^{{(E,j)}}(r)$. The normalization coefficient is found by imposing the condition~\eqref{eq:normalization_equation}, where $C_f$ is related to $\chi^{{(E,j)}}(r_f)$ by {eq.}~\eqref{eq:C_f}. With the wave function normalized, the LDOS can be now calculated using {eqs.}~\eqref{eq:tldos} and~\eqref{eq:ldos}.

Though the assumption that the radial potential $U(r)$ is flat near the origin ($r<r_i$) is really observed in different CGQD-models~\cite{matulis2008quasibound,hewageegana2008electron,recher2009bound,bardarson2009electrostatic,gutierrez2016imaging}, with regard to the confinement potentials as those induced by the STM-tip in experiments reported in Ref.~\cite{zhao2015creating,freitag2016electrostatically,lee2016imaging}, there is a need to relax this assumption. 
Note that, on the other hand, any electrostatic confinement potential should tend to be constant at large distances. 

{When the potential $U(r)$ is not flat near the origin, the wave function~\eqref{eq:flat_initial} is no longer an exact solution of eq.~\eqref{eq:dirac}.} However, if $U(r)$ is continuous near the origin, the asymptotic form of~\eqref{eq:flat_initial} still correctly describes the asymptotic behaviour of the solution, namely  $\chi^{{(E,j)}} (r) \sim ( \propto r^{|j-\frac{1}{2}|}, \propto r^{|j+\frac{1}{2}|})$ [Here, to avoid irrelevant factors we use the symbols $\propto$]. {However, except for $j = \pm \frac{1}{2}$, this asymptotic solution vanishes at the origin ($r=0$), thus cannot be used as the initial value to solve the differential eq.~\eqref{eq:dirac} for $\chi^{{(E,j)}}(r)$}. The simple trick to get around this problem is to {introduce} the regularized wave function $\tilde{\chi}^{{(E,j)}}(r) = r^{-\beta} \chi^{(E,j)} (r)$, where $\beta= \min \{|j-\frac{1}{2}|,|j+\frac{1}{2}|\}$. {This regularized}  wave function $\tilde{\chi}^{{(E,j)}}(r)$ follows the evolution
\begin{equation}
i \frac{\partial \tilde{\chi}^{{(E,j)}}}{\partial r} = \tilde{\mathcal{H}}_r \tilde{\chi}^{{(E,j)}}(r)
\label{eq:dirac_mod}
\end{equation}
with 
\begin{equation}
\tilde{\mathcal{H}}_r = \mathcal{H}_r  - \frac{i\beta}{r}.
\end{equation}
The initial condition for the regularized wave function $\tilde{\chi}^{{(E,j)}}(r)$ is now regular, namely, $\tilde{\chi}^{{(E,j)}} (0)=(\tilde{\mathcal{N}},0)^{T}$ if $|j-\frac{1}{2}| < |j+\frac{1}{2}|$ and $\tilde{\chi}^{{(E,j)}} (0)=(0,\tilde{\mathcal{N}})^{T}$ otherwise. Here $\tilde{\mathcal{N}}$ is the normalization coefficient, which is found using eqs.~\eqref{eq:C_f} and~\eqref{eq:normalization_equation} as described above. With the wave function {$\chi^{(E,j)}(r)$} determined from $\tilde{\chi}^{{(E,j)}}(r)$, the LDOS can again be calculated using eqs.~\eqref{eq:tldos} and~\eqref{eq:ldos}.
 
As a typical illustration for the suggested approach, we calculate the LDOS for the CGQD studied in Ref.~\cite{gutierrez2016imaging}. This CGQD is believed to exhibit a sharp boundary so that the radial confinement potential can be modeled as a step one, $U(r) = V_0 \Theta ({R_0} - r)$, where $V_0 = \operatorname{constant}$, $\Theta (x)$ is the Heaviside step function, and ${R_0}$ is the dot radius. \cnew{In Ref.~\cite{gutierrez2016imaging}, $R_0=5.93$ nm, $V_0= 0.43$ eV and the potential is applied to a piece of graphene with dimension that can be estimated to be $\approx 40$ nm. There it was also shown that this piece can be well treated as an infinite graphene sheet}. For this step potential $U(r)$ the LDOS can be calculated from eqs.~\eqref{eq:tldos} and~\eqref{eq:ldos} with the wave function $\chi^{(j,E)} (r)$ determined directly from eq.~\eqref{eq:dirac}. Note that without an external magnetic field, the energy spectra are identical for $j > 0$ and $j < 0$~\cite{chen2007fock,hewageegana2008electron}, we therefore consider only the case $j > 0$.

\begin{figure}[hbt]
\includegraphics[width=0.42\textwidth]{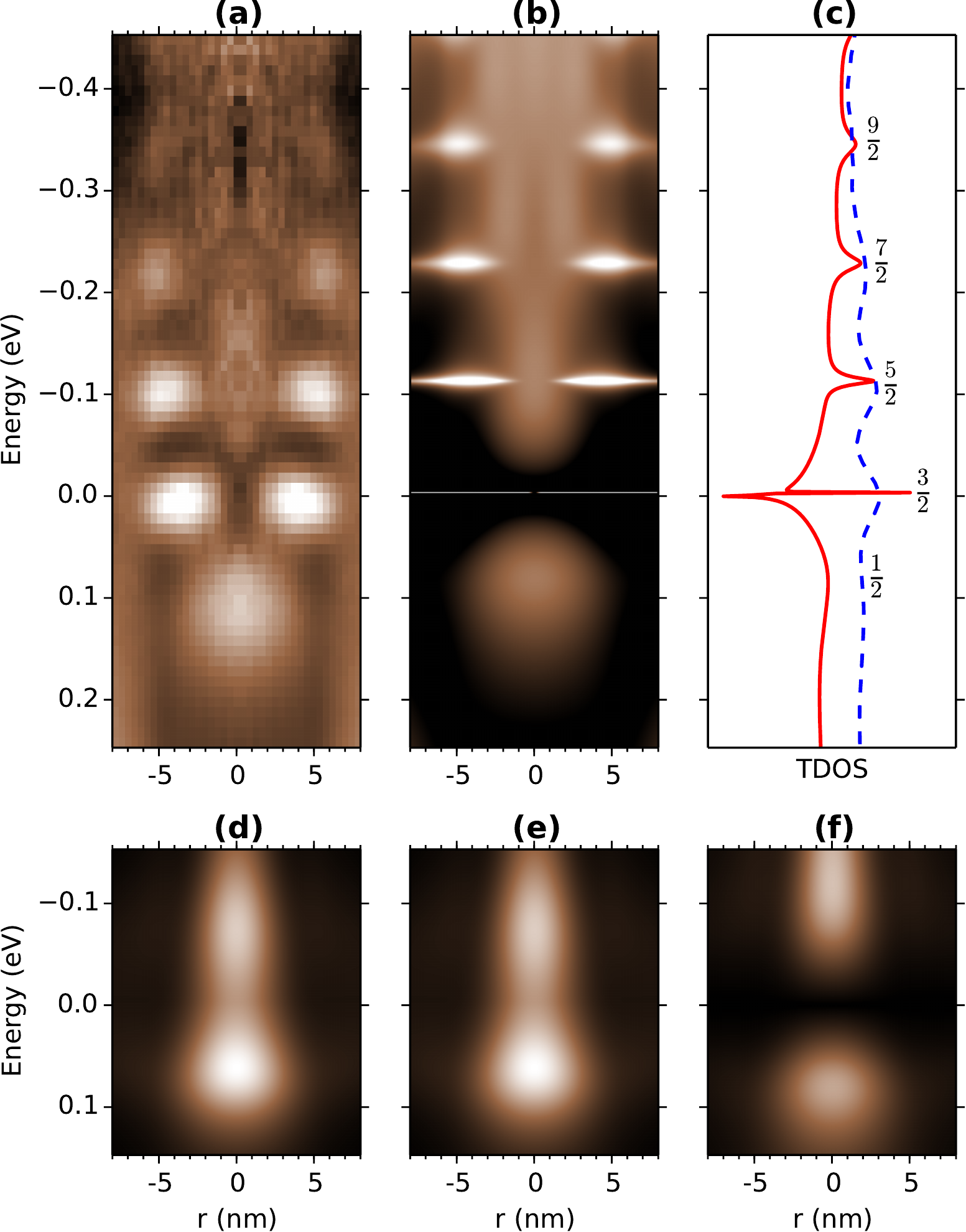}
\caption{(Colour online) LDOSs of CGQD with ${R_0}=5.93$ nm, $V_0= 0.43$ eV ({the background correction of $E_D= -0.347 \mathrm{eV}$~\cite{gutierrez2016imaging} has been subtracted from the raw data leading to a shift in energy when compared to the original plot of Ref.~\cite{gutierrez2016imaging})}: $(a)$ Experimental data provided by the authors of Ref.~\cite{gutierrez2016imaging}; $(b)$ Calculated results using the present approach; $(c)$ Two TDOSs calculated from the data in $(a)$ (dashed) and $(b)$ (solid line) [log scale, arbitrary unit]. The angular momenta $j$ of the resonances are indicated by the nearby numbers. Panels $(d - f)$ compare the partial LDOSs for the state of $j = \frac{1}{2}$: $(d)$ from Ref.~\cite{gutierrez2016imaging}; $(e)$ Eq.~\eqref{eq:ldos} without normalization; and $(f)$ Eq.~\eqref{eq:ldos} with normalization.}
\label{fig:1}
\end{figure} 

{In Fig.~\ref{fig:1}$(a)$ and $(b)$, we show respectively the experimental data provided by the authors of Ref.~\cite{gutierrez2016imaging} and the LDOS calculated using our approach for the same CGQD}. Both figures were plotted in the same format, giving a clear view of the full spectrum through a cross-sectional slice of the CGQD. {To access to the resonances in the LDOS spectra, we calculate the corresponding experimental and theoretical total density of states (TDOS) via
\begin{equation}
\rho (E) = \int_0^{R_\mathrm{max}} 4 \pi \operatorname{d} r \sum_{j=\frac{1}{2}}^{+\infty}\rho^{(j)} (E,r),
\label{eq:tdos}
\end{equation}	
where $R_\mathrm{max}$ is the maximal radius probed in the experiment (here $R_\mathrm{max}=8 \ \mathrm{nm}$), which encircles the major maxima of the LDOS. 
For the theoretical LDOS, the summation over angular momenta is truncated at $j_\mathrm{max}= \frac{31}{2}$ as higher momenta do not contribute significantly to the LDOS at this energy scale. \cnew{The obtained theoretical and experimental LDOSs are presented in Fig.~\ref{fig:1}(c) by the solid and dashed lines, respectively.}} 

{Comparing Fig~\ref{fig:1}(a) and Fig~\ref{fig:1}(b),} obviously, on the whole, there is a very good agreement on the resonance energies between the experimental data and our theoretical calculation. Particularly, our calculation gives a better agreement for the state of lowest angular momentum $j=\frac{1}{2}$, compared to the calculation reported in Ref.~\cite{gutierrez2016imaging}. A closer analysis shows that the difference between the two calculated results mainly stems from the normalization eq.~\eqref{eq:normalization_equation}. This normalization was absent from the scattering calculations in Ref.~\cite{gutierrez2016imaging}, but naturally appears in the direct calculation described above. As a particular verification of this intuitive assessment, we compare the partial LDOSs for just the state of $j = \frac{1}{2}$ those calculated: from the scattering approach in Ref.~\cite{gutierrez2016imaging} [Fig.~\ref{fig:1}$(d)$]; from eq.~\eqref{eq:ldos} discarding the normalization~\eqref{eq:normalization_equation} [Fig.~\ref{fig:1}$(e)$]; and from eq.~\eqref{eq:ldos} with the  normalization~\eqref{eq:normalization_equation} [Fig.~\ref{fig:1}$(f)$]. Obviously, while the two figures Fig.~\ref{fig:1}$(d)$ and Fig.~\ref{fig:1}$(e)$ are practically identical, the normalization pushes the state in Fig.~\ref{fig:1}$(f)$ closer to the experimental position. It should be noted that, despite this improvement, a small discrepancy between the theoretical prediction and the experimental position of this state still persists. We speculate that, having the largest level width{~\cite{chen2007fock,matulis2008quasibound}}, the $j = \frac{1}{2}$ state is more susceptible to various fluctuations such as the imperfection of the dot boundary (as suggested in Ref.~\cite{gutierrez2016imaging}) or the thermal noise, which further affect its position.

{Another level that is also particularly interesting is that of $j=\frac{3}{2}$. It shows very small level width in comparison to the other levels. With very long wave length and energy near the zero point, it is in fact resembles the so-called zero-energy bound state~\cite{downing2011zero}.}

\begin{figure}[!bth]
	\includegraphics[width=0.45\textwidth]{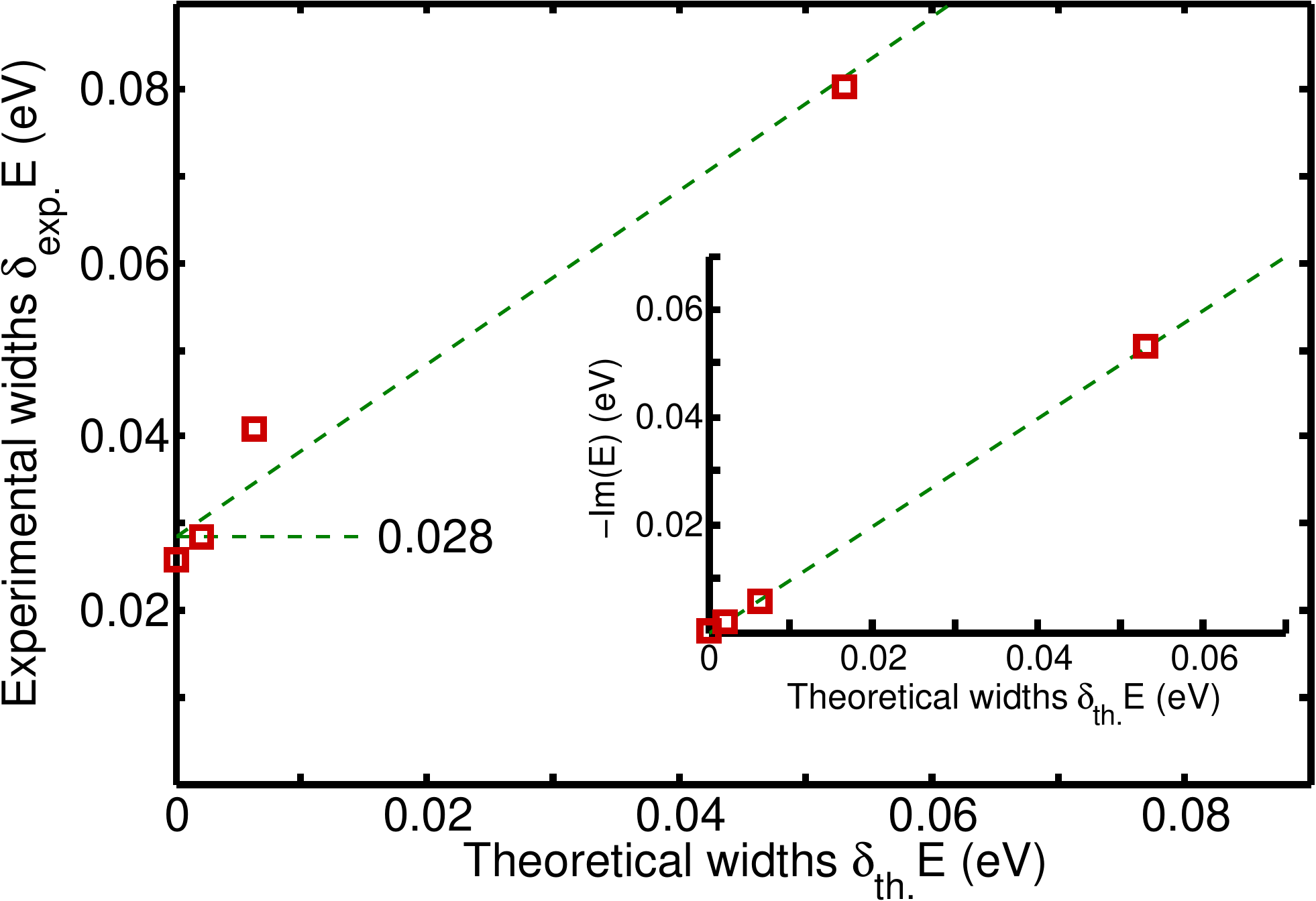}
	\caption{(Colour online) Resonance widths of the QBSs of $j = \frac{3}{2}$, $\frac{5}{2}$, $\frac{7}{2}$, and $\frac{1}{2}$ (from left to right in both main figure and inset) indicated in Fig.~\ref{fig:1}$(c)$. Main figure: Comparison of resonance widths extracted from the experimental ($\delta_{\mathrm{exp.}}E$) and theoretical ($\delta_{\mathrm{th.}}E$) TDOSs in Fig.~\ref{fig:1}$(c)$: vertical axis - experimental widths and horizontal - theoretical ones. The dashed straight-line with unity slop and the fitted offset of $0.028\ \mathrm{eV}$ is set to show a systematically experimental-theoretical discrepancy. Inset: the widths extracted from calculated TDOS and corresponding quantities $-\Im E$ are in comparison. The dashed straight-line has unity slop and zero offset.}
	\label{fig:2}
\end{figure}

Although the experimental and theoretical TDOS-lines in Fig.~\ref{fig:1}{$(c)$} show the resonances at almost the same energies, we also notice that the theoretical widths are noticeably smaller than the experimental ones. To assess this experimental-theoretical discrepancy, we determine the widths of both the experimental \cnew{($\delta_{\mathrm{exp.}}E$)} and theoretical \cnew{($\delta_{\mathrm{th.}}E$)} resonances in Fig.~\ref{fig:1}{$(c)$} by fitting the TDOS around each resonance to a Lorentzian peak~\cite{matulis2008quasibound, gutierrez2016imaging}. The resonance of $j=\frac{9}{2}$ is excluded due to the low quality of the experimental data. For the rest, the widths obtained from the two TDOSs are compared in Fig.~\ref{fig:2}. It seems that for all the resonances examined, the experimental widths (vertical axis) are in the same amount of $\approx 0.028\ \mathrm{eV}$ larger than the theoretical ones (horizontal axis). This systematic smearing of resonances (that makes peaks wider and lower) may be caused by, as was already noted in Ref.~\cite{gutierrez2016imaging}, the fact that the electrons in graphene have a non-zero probability of transition into the surface of the copper substrate, which leads to a decrease of the trapping times in QBSs. The thermal noise might be an additional reason for this resonance smearing.

For the same QBSs of $j = \frac{1}{2}$, $\frac{3}{2}$, $\frac{5}{2}$, and $\frac{7}{2}$ of the studied CGQD, we also calculate $(-\Im E)$ of the QBS complex energies $E$ using our $T$-matrix approach suggested in Ref.~\cite{nguyen2016transfer}. For a given QBS, the quantity $(- \Im E)$ should provide a direct measure of the resonance width. In the inset in Fig.~\ref{fig:2}, we show the obtained $(- \Im E)$ in comparison with the resonance widths extracted from the calculated TDOS. We find that for all the resonances examined, the resonance widths extracted from TDOS (horizontal axis) and the corresponding imaginary parts of the QBS complex energies (vertical axis)  {are in agreement with the relative accuracy of at least $92\%$}. This gives an additional confidence to the current discussion. 

\begin{figure}[thb]
\includegraphics[width=0.45\textwidth]{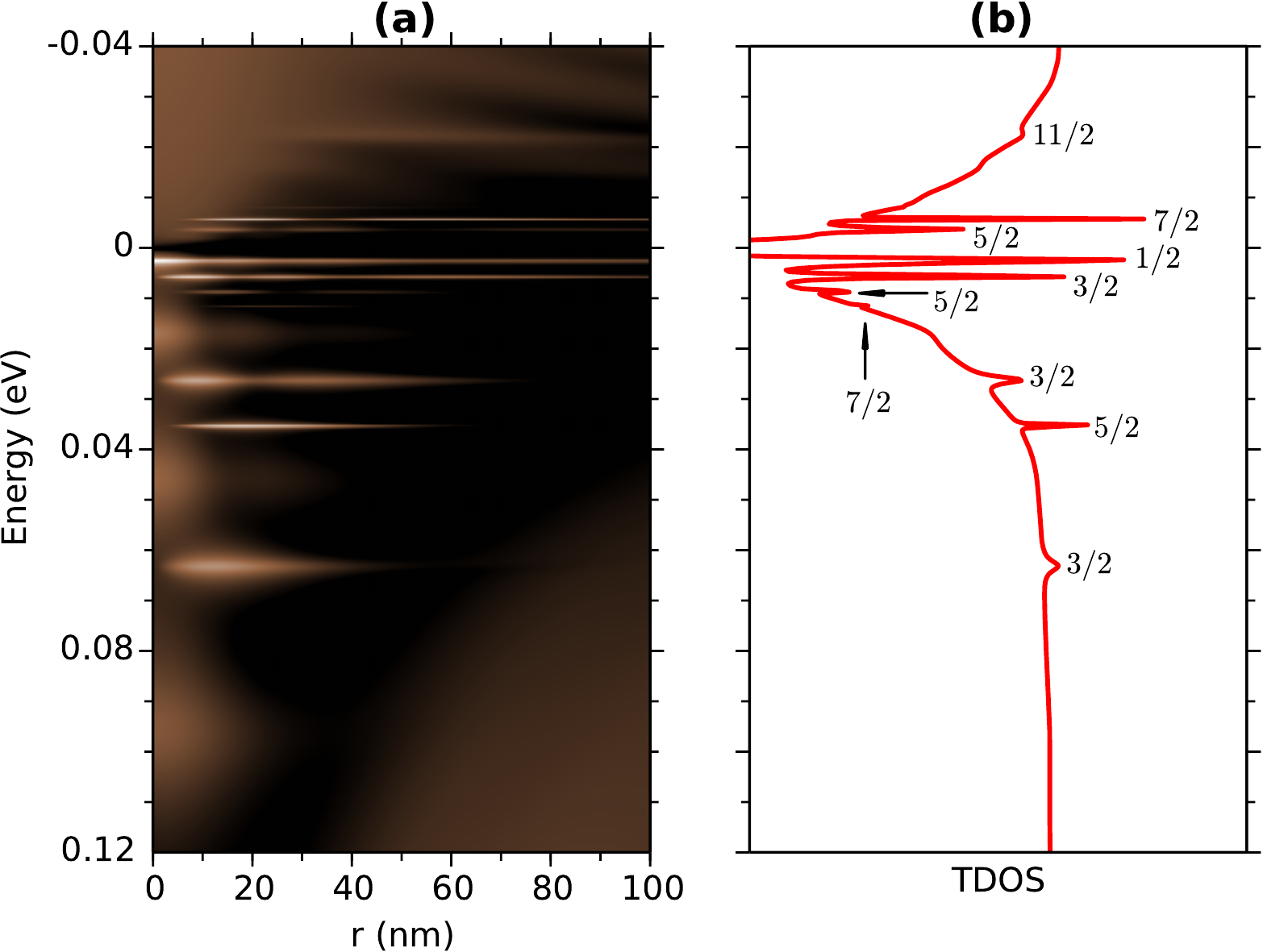}
\caption{(Colour online) LDOS $(a)$ and TDOS $(b)$ are presented in the way similar to Fig.1, but for the CGQD created by the Lorentzian potential with $V_0= 0.2$ eV and  $\bar{R}= 30$ nm. The angular momenta $j$ of the major resonances are indicated by the nearby numbers. \cnew{Note that the dip at $E \approx 0$ of the TDOS reflects the vanishing density of states at the Dirac point in the pristine graphene.}}
\label{fig:3}
\end{figure}

As another illustration, we calculate the LDOS for the CGQDs of the type that is created by a charged STM-tip in the experiments reported in Refs.~\cite{zhao2015creating,lee2016imaging}. {For such the CGQDs we follow Ref.~\cite{downing2011zero} and model the smooth confinement potentials as:
$U(r) = V_0 / [ 1 + (r / \bar{R} )^2 ]$ (Lorentzian potentials),
where $V_0$ and $
\bar{R}$ measure the strength and the width of the potential, respectively}. Although the LDOS can still be calculated from eqs.~\eqref{eq:tldos} and~\eqref{eq:ldos}, since the potential is not flat in the vicinity of the origin, the wave function $\chi^{(E,j)}(r)$ in these equations should be determined from the regularized one $\tilde{\chi}^{{(E,j)}}(r)$ of eq.~\eqref{eq:dirac_mod}. 
\cnew{We use the \texttt{ODE45}~\cite{Shampine1997} solver to solve this differential equation (with the maximal step size $10^{-2} \bar{R}$). Note that the differential equation for $\tilde{\chi}$ is solved successively from $r_i=10^{-8} \bar{R}$ towards infinity; so there is no approximation in this step when one stops the solver at some $r=r_f$. In the present LDOS calculation, we stop the solver at $r_f=10 \bar{R}$ to calculate $C_f$ from eq.~\eqref{eq:C_f}. This is equivalent to (approximately) regarding $U(r)=0$ for $r \ge r_f= 10 \bar{R}$.}
The TDOS is calculated by integrating eq.~\eqref{eq:tdos} in the same area, $R_{\mathrm{max}}=10 \bar{R}$ (with the angular momentum series also terminated at $j_{\mathrm{max}}=\frac{31}{2}$). We show in Fig.~\ref{fig:3} the LDOS $(a)$ and the corresponding TDOS $(b)$ calculated for the CGQD with the Lorentzian confinement potential of $\bar{R} = 30 \ \mathrm{nm}$ and $V_0 = 0.2$ eV. Certainly, from this TDOS, we can extract the resonance widths in the same way as presented above. Owing to the lack of detailed experimental data available for comparison, we would like simply to note the rather dense and narrow resonances emerged in Fig.~\ref{fig:3}(b). The whole spectrum is also very sensitive to both parameters $V_0$ and $\bar{R}$.   

Thus we have presented an approach to calculate the LDOS of CGQDs. This approach equally applies to practically any structure created by axially symmetric electrostatic potentials on a continuous graphene sheet. {It can be easily extended to include a mass term in the Hamiltonian~\eqref{eq:hamilton}~\cite{nguyen2016transfer}}. {Under an external magnetic field, the current formulation does not however apply directly and further studies are needed; see~\cite{recher2009bound,Givaras2011a,Givaras2012a,kim2012resonant} for alternative approaches}.

{\sl Acknowledgement:} We would like to thank Abhay Narayan Pasupathy and Christopher Gutierrez for sharing their excellent experimental data and for helpful discussions. We also thank C. Huy Pham for his comments on the draft. This research is funded by Vietnam National Foundation for Science and Technology Development (NAFOSTED) under grant number 103.02-2015.48.

\bibliography{chau04}
\end{document}